\documentclass[10pt,a4paper]{article}

    \usepackage{inputenc}
    \usepackage[english]{babel}
\usepackage[margin=1in]{geometry}

\usepackage[compact,raggedright,small]{titlesec}
\usepackage[affil-it]{authblk}
 \usepackage[runin]{abstract}
\usepackage{multicol}
\usepackage[dvips]{graphicx}

    \addto{\Affilfont}{\small}

    \title{\large\bfseries Parity violating effects in an exotic perturbation of the  rigid rotator}
    \date{\small (March 09, 2015)}
    \author[1]{A.\ Pallares-Rivera\footnote{\texttt{pallares@ifisica.uaslp.mx}}}
    \author[1]{M.\ Kirchbach}
    \affil[1]{Institute of Physics, Autonomous University at San Luis Potosi,\par
Av. Manuel Nava 6, University Campus, San Luis Potosi, SLP 78290, Mexico}

\begin{document}

\maketitle

 \begin{abstract}
The perturbation of the free rigid rotator by the trigonometric Scarf potential is shown to conserve its 
 energy excitation patterns and change only the wave functions towards  spherical harmonics 
 rescaled by a function of an unspecified parity, or 
 mixtures of such rescaled harmonics of equal magnetic quantum numbers and different angular momenta. In effect, 
 no parity can be assigned to the states of the rotational bands emerging in this exotic way, 
 and the electric dipole operator is allowed to acquire non-vanishing expectation values.
      \end{abstract}

\begin{flushleft}
{\bf Keywords:} rigid rotator, trigonometric Scarf potential,  
degeneracy conservation, parity violation, non-vanishing electric dipole moment
\end{flushleft}

 
 \section{Introduction} 
 
\begin{figure}[H]
\centering
\begin{tabular}{cc}
\resizebox{2.5cm}{!}{\includegraphics{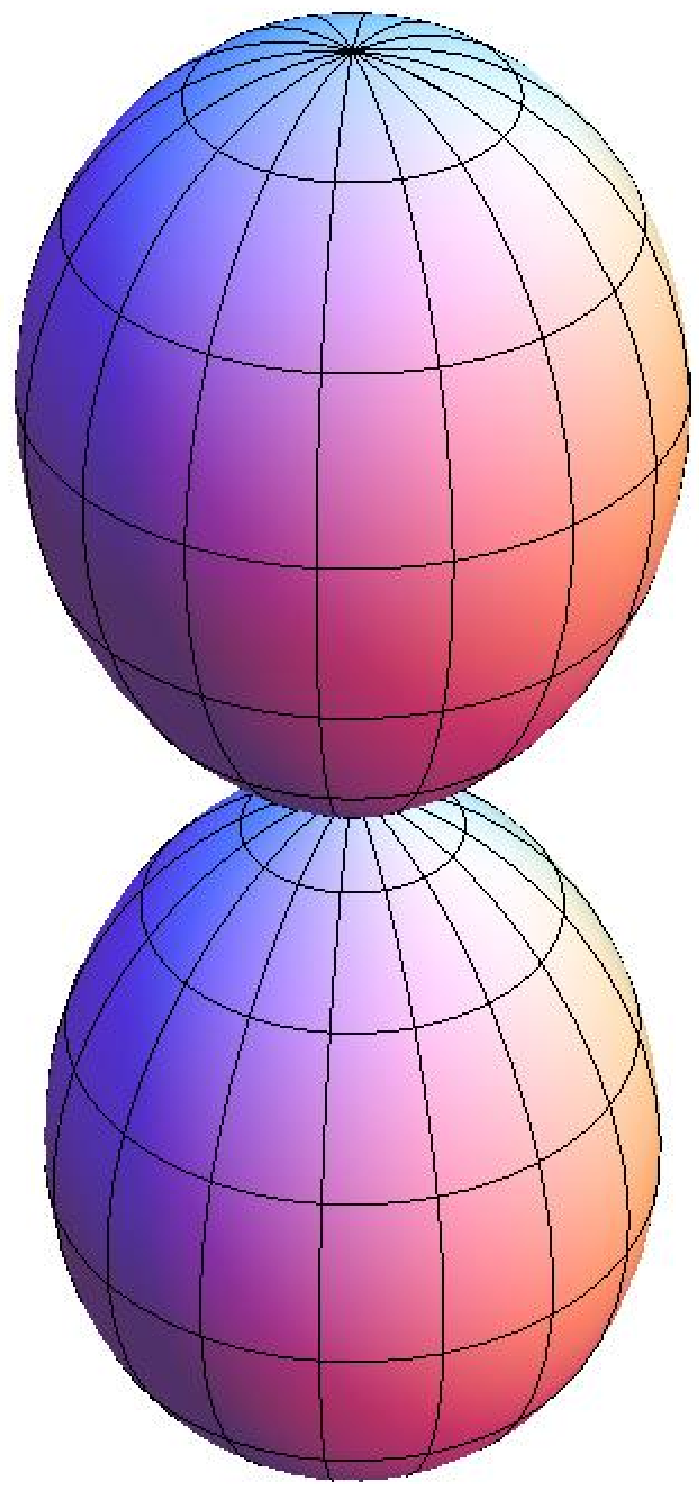}} & \resizebox{5cm}{!}{\includegraphics{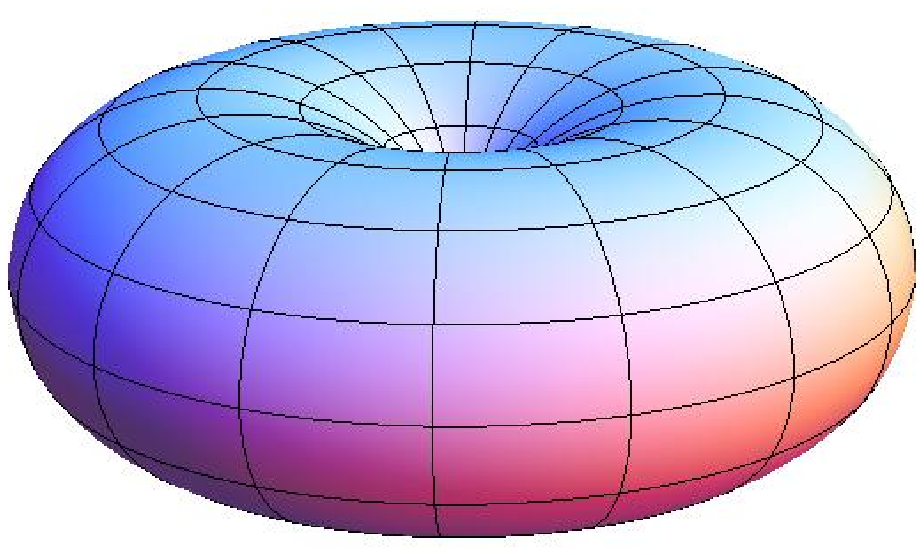}} \\
(a) $|Y_{1}^{0}\left(\theta,\varphi\right))|^2$ & (b) $|Y_{1}^{\pm 1}\left(\theta,\varphi\right)|^2$
\end{tabular}
\caption{The three different  shapes of the probability density distributions according to which
one and the same energy, like $E_{J=1}=2{\mathcal B}$,  corresponding to the first excited level,  is released in the
$|1M>\to |00>$ ground state transition, \textit{i.e.} spherical polar plots
of  the $|Y_1^{0}\left(\theta,\varphi\right)|^2$ and 
$|Y_1^{\pm 1}\left(\theta,\varphi\right)|^2$ functions, defined in (\ref{Gl4}). The right figure counts twice because the radiation can go clockwise
or, counter-clockwise.}
\label{Fig:1}       
\end{figure}

 The energy spectra of molecules consisting of two atoms are as a rule well understood in terms of equidistant vibrational 
 excitations characterized by wavelengths in the millimetric infrared range, on top of which one observes 
 rotational bands of energies growing as $J (J+1)$, with $J$ non-negative integer,  which disintegrate by emitting 
 nanometric microwave radiations (so called rovibron model \cite{Harris}). 
 {}For a fixed vibrational mode, the microwave radiation  is entirely attributed to 
 the proportionality of the molecular Hamiltonian  to the squared angular  momentum operator, 
 ${\mathbf J}^2$, 
 \begin{equation} 
 {\mathcal H}_{rot}(\theta,\varphi)=\frac{\hbar^2}{2\mu{R}_e^2}{\mathbf J}^2(\theta,\varphi), 
 \quad{\mathbf J}^2(\theta,\varphi)=-\frac{1}{\sin\theta}\frac{\partial }{\partial \theta}\sin\theta \frac{\partial }{\partial \theta} 
 -\frac{\frac{\partial^2}{\partial \varphi 2}}{\sin^2\theta}, 
 \label{rtr_Ham} 
 \end{equation} 
 where $\mu$ stands for the reduced mass of the two molecules, and $R_e$ denotes the equilibrium bound length. 
 The description of the rotational modes amounts to the diagonalizing of ${\mathcal H}_{rot}(\theta,\varphi)$ in the basis of the  functions, 
 $\Phi(\theta,\varphi)$ of the polar, $\theta$, and azimuthal, $\varphi$, angles, 
 \begin{eqnarray} 
 {\mathcal H}_{rot}(\theta,\varphi)\Phi  (\theta,\varphi)&=&E\Phi(\theta,\varphi). 
 \label{Gl2} 
 \end{eqnarray} 
 Interpreting $\mu R_e^2$ as the  moment of inertia, ${\mathcal I}$, about the center of mass, 
 the equation ~(\ref{Gl2}) is cast in the standard form of the linear rigid rotor, termed to here  as 
 rigid  rotator, according to, 
 \begin{eqnarray} 
 {\mathcal B}{\mathbf J}^2(\theta,\varphi)Y_J^M(\theta,\varphi)&=&E_JY_J^M (\theta,\varphi),\quad 
 E_J= {\mathcal B}J(J+1), \quad {\mathcal B}=\frac{\hbar^2}{2{\mathcal I}}, 
 \label{Gl3} 
 \end{eqnarray} 
 with ${\mathcal B}$ being the rotational constant. 
 The solutions to the latter equation are the well known three-dimensional  spherical harmonics, $Y_J^M(\theta,\varphi)$, i.e. 
 $\Phi(\theta,\varphi)=Y_J^M(\theta,\varphi)$, defined as, 
 \begin{equation} 
 Y_J^M(\theta, \varphi)=e^{iM\varphi} P_J ^M(\cos\theta), \quad J=0,1,2,... 
 \label{Gl4} 
 \end{equation} 
 Here, $P_J^M(\cos\theta)$ are the associated Legendre functions \cite{Arfken}, known to  become the Legendre polynomials, 
 $P_J(\cos\theta)$,  for $M=0$. 
 It is of common use to cast the rotational spectrum  in (\ref{Gl3}) in terms of frequencies, $\nu(J)$,  as 
 \begin{equation} 
 \nu (J)=\frac{1}{h}E_{{rot}}=\frac{1}{h} \left( E_J-E_{(J-1)}\right)= 
 2\frac{{\mathcal B}}{h}J, 
 \label{freq} 
 \end{equation} 

 and  define spectral lines separated by the constant gap of $2{\mathcal B}/h$. 
 The above considerations refer to the idealized case of a pure rotator,  the reality being that  the equidistant line spacings are altered by 
 centrifugal distortions and other secondary effects, left aside in what follows. 
 The rotational spectra are remarkable through the fact that the energy, $E_J$, equivalently, the frequency, $\nu(J)$, 
 in (\ref{freq}),  depends on $J$ alone meaning that 
 the multiplicity of states in a level is $(2J+1)$-fold. Also the shapes 
 of the disintegration (``radiation'') patterns, more precisely, the angular probability density distributions, 
 given by the spherical polar plots of the 
 squared $Y_J^M(\theta,\varphi)$ functions, are energy degenerate. The plot in fig. \ref{Fig:1} illustrates 
 the degeneracy of the probability density  distributions responsible for  the disintegration of the states 
 $|J=1, M=0,\pm 1>$ to the ground state. 
 
 Our principal goal is to answer the question whether the shapes of the probability density distributions  are uniquely 
 determined by the spectral frequencies, which parallels  the celebrated question asked by  Mark Kac in 1966, 
 on whether the sound frequencies of a drum specify its form in an unique way \cite{MKac}. 
 Our case is that similarly to Kac's question, also the answer to the one posed here by us turns out to be negative. 
 
 {Our point is that the perturbation of ${\mathcal H}_{rot}(\theta,\varphi)$ by a properly designed 
 potential can retain the spectral frequencies  and alter the wave functions. 
 In this fashion, two sets of identical isospectral frequency patterns will correspond to  two  sets of probability density distributions of non-equivalent shapes. 
 Stated differently, according to our findings,  $(2J +1)$-fold degeneracies do not necessarily imply equality of the Hamiltonian 
 neither to the one  of the canonical rigid rotator, nor to a similarity transformation of it. 
 To be specific, below we show that the energies in the spectrum of the rigid rotator, perturbed by the 
 trigonometric Scarf potential, 
 remain identical to  those within the spectrum of the free rotator,  though the wave functions of the former are of unspecified parities and quite distinct 
 from those of the latter,  which are the spherical harmonics of well defined parities. 
 The consequence will be rotational levels of non-vanishing electric dipole moments.} 
 The paper is structured as follows. In the next section we place the goal of our investigation within the context of spectral geometry 
 and isospectral potentials. In section 3  we introduce  the perturbation of the rigid rotator by the trigonometric Scarf potential and  present its spectrum and wave functions. 
 In section 4 we analyze the wave functions emerging  after the perturbation and show they have lost their spatial parities, thus allowing for non-vanishing electric dipole expectation values. 
 The paper closes with brief conclusions. 
 
 \section{Isospectral potentials from the perspective of spectral geometry} 

Spectral geometry studies the relation between spectral properties of 
elliptic differential or pseudo-differentials operators 
and the associated Riemannian manifolds, among them curved surfaces. 
A Riemannian manifold, $(M, g)$,  is characterized by its metric $g$ with covariant components, 
$g_{\alpha \beta}$, and its Laplacian 
which is a second order differential operator acting on symmetric 
traceless tensors of second ranks, 
viewed as infinitesimal metric transformations. The  general definition of 
a Laplacian in the spirit of Lichnerowicz reads, 
${\mathbf \Delta}_g=-\mbox{div}(\nabla)= -\nabla ^\alpha \nabla_\alpha 
=-g^{\beta\gamma}\nabla_\beta\nabla_\gamma$, 
with the minus sign ensuring a spectrum bounded from below.

 A further key characteristic of a curved surface, in the following restricted to a compact Riemannian space, 
 is the eigenvalue problem of the Laplace operator, ${\mathbf \Delta}_g$ on this metric, 
 \begin{equation} 
 {\mathbf \Delta}_g  f_k =\lambda_k f_k, \quad i=0,1,2,....   
 \label{IP_1} 
 \end{equation} 
 with $f_k$ being the corresponding eigenfunctions. The discrete number sequence, 
 $(0=)\lambda_0\leq \lambda_1\leq \lambda_2\leq ... \to \infty $ 
 is referred to as the spectrum of ${\mathbf \Delta}_g$, or,  of $M$. 
 The vibrations of a thin membrane (a drum) are fixed by the eigenfunctions 
 of the relevant Laplacian with Dirichlet boundary conditions and the square roots, $\sqrt{\lambda_i}$, 
 of the eigenvalues are proportional to the sound frequencies. 
 The fundamental questions asked by mathematicians in spectral geometries (see \cite{VitoCruz} for a pedagogical text) 
 concern the properties of the spectrum given the geometry 
 (the metric)  and vice verse, whether the geometry (the metric) can be restored 
 given the spectrum. While the first question is straightforwardly answered by solving the  eigenvalue problem of the Laplacian, the 
 answer to the inverse spectrum problem is far from being obvious. 
 The issue is that the  information provided by the spectrum alone is insufficient to allow one to uniquely 
 recover the  metric of the surface. Mark Kac's question mentioned in the introduction on   
 whether one can hear the shape of a drum \cite{MKac} 
 expresses precisely this very  non-uniqueness of the solutions of the inverse spectral problem. 
 Indeed, several drums of strictly equal spectra and non-equivalent shapes have been constructed 
 beginning with  work by Gordon and Wilson in 1984 \cite{Carolyn}. 
 The applications of spectral geometry to physics  are highly interesting and allow one to handle 
 the so called isospectral potential problems \cite{Dorothee}. The latter refer to two flat space Schr\"odinger equations with distinct potentials, 
 $V_1$ and $V_2$, 
 \begin{eqnarray} 
 H^{(1)}\psi_i^{(1)}&=&\left( 
 -\frac{\hbar^2}{2M}\nabla^2 +V_1\right)\psi_i^{(1)}=E^{(1)}_i\psi_i^{(1)}, \nonumber\\ 
 H^{(2)}\psi_i^{(2)}&=&\left(-\frac{\hbar^2}{2M}\nabla^2 +V_2\right)\psi_i^{(2)}=E_i^{(2)}\psi_i^{(2)},\nonumber\\ 
 E_i^{(1)}&=&E_i^{(2)}\equiv E_i, 
 \label{IP_2} 
 \end{eqnarray} 
 which nonetheless describe exactly same spectra. Here, $M$ denotes the mass of the particle under consideration. 
 Several such potential pairs  are known in quantum mechanics, the Morse and the hyperbolic 
 Scarf potentials being one of them (see  \cite{Khare} for details). 
 In one of the possibilities, though not necessarily, two isospectral potentials can be related by a so called isospectral transformation which is no more than a  similarity transformation of any one of the  Hamiltonians, 
 say $H^{(1)}$, by some properly selected invertible smooth function, ${\mathbf F}$, according to 
 \begin{eqnarray} 
 H^{(2)}\psi_i^{(2)}=\left[ {\mathbf F}H^{(1)}{\mathbf F}^{-1}\right] \left[ {\mathbf F}\psi_i^{(1)}\right] &=& 
 E_i\left[{\mathbf F}\psi^{(1)}_i\right], 
 \label{IIP_2b} 
 \end{eqnarray} 
 meaning 
 \begin{equation} 
 V_2=\left[ {\mathbf F} H^{(1)}{\mathbf F}^{-1}\right] +\frac{\hbar^2}{2M}\nabla^2, \quad \psi^{(2)}_i= 
 {\mathbf F}\psi^{(1)}_i. 
 \label{IP_2c} 
 \end{equation} 
 In general, potentials generated in this manner are complicated and  contain gradient terms. 
 The link of the Schr\"odinger equation to geometry (so called ``geometrization'' of Schr\"odinger's equation \cite{Karamatskou}) 
 is established on the basis of the observation that the equations in (\ref{IP_2})  equivalently rewrite to eigenvalue problems of 
 Laplacians in the so-called Maupertuis-Jacobi metrics, here denoted by $g^{(MJ)}_{ij}(V_k)$ and defined as, 
 \begin{equation} 
 g^{(MJ)}_{ij}(V_k)=\Omega^2 \delta_{ij}, \quad \Omega^2=2M(E_i-V_k), \quad k=1,2, 
 \label{IP_3} 
 \end{equation} 
 as 
 
 \begin{eqnarray} 
 \left( \hbar^2\widetilde{\nabla^2}_{g^{(MJ)}(V_k)} +1\right)\psi_i^{(1)}&=&0 \nonumber\\ 
 \widetilde{\nabla^2}_{g^{(MJ)}(V_k)}= \Omega^{-2}\nabla^2. 
 \label{IP_4} 
 \end{eqnarray} 
 Here, $\Omega^2$ is the so called ``conformal'' factor. Correspondingly, the  metrics 
 $g^{(MJ)}_{ij}(V_k)$  are referred to as conformal, while   
 $\widetilde {\nabla^2}_{g^{(MJ)}(V_k)}$ are the conformal  Laplacians on the 
 surfaces with the Maupertuis -Jacobi metrics, $g^{(MJ)}(V_k)$ (with $k=1,2$). 
 It should be obvious that the equations in (\ref{IP_4}) refer to isospectral problems on surfaces of distinct metrics 
 and can be treated with the methods of spectral geometry,  as indicated in the opening of this section. 
 
 Back to the inverse spectral problem, 
 in order to restore its uniqueness, new information needs  to be invoked. 
 As such,  mathematicians consider  the numbers of nodes and the nodal lines of the eigenfunctions \cite{NodalLines}. 
 Within the context of spectral geometry the goal of the present study can be cast in the following way: 
 \begin{quote} 
 We draw attention to the fact that the angular  potentials 
 \begin{equation} 
 V_1(\theta)= \frac{M^2 -\frac{1}{4}}{\cos^2\theta}, \quad 
 V_2(\theta) = \frac{M^2 -\frac{1}{4}+b^2}{\cos^2\theta}-\frac{2bM\tan\theta}{\cos\theta} -\frac{1}{4}, 
 \label{IP_5} 
 \end{equation} 
 are isospectral. 
 \end{quote} 
 However, rather than explicitly constructing the respective Maupertuis-Jacobi metrics, 
 we here, being focused on physical applications in general, and on molecular physics in  particular, 
 prefer to visualize the difference of the aforementioned metrics  indirectly.   
 In first place the distinction is made through  the evidently different shapes of the  $V_1(\theta)$ and $V_2(\theta)$ 
 probability density distributions corresponding to equal quantum numbers and same energy, 
 and in second place through the unspecified parities of the  $V_2(\theta)$ eigenfunctions, in contrast to   
 the well defined parities of the $V_1(\theta)$ eigenfunctions.  These  are 
 all physical entities of high spectroscopic relevance and are of major interest 
 to experimental studies of the disintegration modes.\\ 
 
 \noindent 
 Another interesting question regarding isospectral potentials concerns their symmetry properties. 
 As we shall see in the following, the $V_1(\theta)$ potential is rotationally invariant because in properly chosen coordinates, 
 it can be viewed as part of the Laplace operator on the two-dimensional spherical surface, $S^2$. 
 The symmetry properties of the second potential  are not instantly obvious and need to be elaborated separately. 
 {}For this purpose one first has to systematically search for a complete set of operators which commute with 
 $\widetilde{\nabla^2}_{g^{(MJ)}(V_2)}$ and and then investigate their commutators with the aim to possibly figure out a 
 relevant Lie algebra. This question, in going far beyond the scope of our investigation,  will be left aside for the 
 time being, and will  be attended properly in a future work.   
 The next section is devoted to the presentation of our case.   
 
 \section{Exotic degeneracy preserving  perturbation of the rigid rotator } 
 {}In the current section we choose to simplify  notations for the sake of transparency of the formulas 
 and  work  in dimensionless units, setting ${\mathcal B}=1$. 
 In eq.~ (\ref{Gl2}) the  parametrization  of the two dimensional sphere is such that the polar angle $\theta$ 
 has been read off from the positive $z$ axis (the North pole). 
 Instead, we here  take the liberty (without any loss of generality) to read off this angle from the negative $y$-axis, 
 meaning that instead of tracing  meridians in the usual way along  North-East-South-West-North direction, we  trace them 
 along West-North-East-South-West. The reason behind this re-parametrization of the sphere will be explained shortly below, after 
 the equation (\ref{S2_1DSchr}).   
 In so doing, the expression for ${\mathbf J}^2$ and the spherical harmonics 
 correspondingly changes to, 
 \begin{eqnarray} 
 {\mathbf J}^2(\theta,\varphi)=-\frac{1}{\cos\theta}\frac{\partial }{\partial \theta }\cos\theta 
 \frac{\partial}{\partial \theta} +\frac{J_z^2}{\cos^2\theta}= -\frac{\partial ^2}{\partial\theta^2} 
 +2\tan \theta \frac{\partial }{\partial \theta}  - \frac{M^2}{\cos^2\theta },&& 
 \label{L2ang}\\ 
 {\mathbf J}^2(\theta,\varphi)Y_{J}^M(\theta,\varphi) 
 =J(J+1)Y_{J}^M(\theta,\varphi),&& 
 \label{L2eprblm} 
 \\ 
 J_z(\varphi) =-i\frac{\partial}{\partial\varphi}, \quad Y_{J}^M(\theta,\varphi)=e^{iM\varphi}P_{J}^M(\sin\theta).&& 
 \label{Ylm} 
 \end{eqnarray} 
 With that in mind, we now consider the  perturbation of the rigid-rotator by the following potential, 
 ${\mathcal V}(\theta)$, 
 \begin{eqnarray} 
 {\mathbf J}^2(\theta,\varphi)&\to& H(\theta,\varphi) ={\mathbf J}^2(\theta,\varphi)+ 
 {\mathcal V}(\theta),\nonumber\\ 
 {\mathcal V}(\theta)&=&\frac{b^2 }{\cos^2\theta} 
 -\frac{2b M\tan \theta  }{\cos\theta } -\frac{1}{4}. 
 \label{Pert_Sph} 
 \end{eqnarray} 
 The eigenvalue problem of the perturbed Hamiltonian now takes the form, 
 \begin{eqnarray} 
 H(\theta,\varphi) \Psi (\theta,\varphi) =\epsilon\Psi (\theta, \varphi), \,\,\Psi(\theta,\varphi)&=& 
 \phi(\theta)e^{i{M}\varphi}. 
 \end{eqnarray} 
 The second order operator (\ref{L2ang}) contains a first order derivative, 
 which can be eliminated under the following  variable change, 
 \begin{equation} 
 U(\theta)=\sqrt{\cos\theta}\phi(\theta). 
 \label{var_chng} 
 \end{equation} 
 In effect, (\ref{Pert_Sph}) acquires the shape of  the following one-dimensional (1D) Schr\"odinger equation in the $\theta$ variable, 
 \begin{eqnarray} 
 {\mathcal H}(\theta)U (\theta)=\epsilon U(\theta),\quad 
 {\mathcal H}(\theta)=\left[-\frac{{\mathrm d}^2}{{\mathrm d}\theta^2}+V_{ScI}(\theta)\right],&& 
 \label{Pert_Schr} 
 \end{eqnarray} 
 with the Schr\"odinger perturbation potential, here denoted by $V_{ScI}(\theta)$,  being  given as, 
 
 \begin{eqnarray} 
 V_{ScI}(\theta)&=&\frac{b^2 +a(a+1)}{\cos^2\theta} 
 -\frac{b\left(2a +1 \right)\tan \theta  }{\cos\theta } 
 -\frac{1}{4},\nonumber\\ 
 a&=&|M|-\frac{1}{2}. 
 \label{ScfI} 
 \end{eqnarray} 
 The squared  magnetic quantum number, $M^2$,  in the operator on the sphere in (\ref{L2ang}), 
 gets replaced in the one-dimensional Schr\"odinger operator by, 
 
 \begin{equation} 
 M^2\longrightarrow M^2-\frac{1}{4}=a(a+1)=\left(|M|-\frac{1}{2}\right)\left(|M|+\frac{1}{2}\right),   
 \label{S2_1DSchr} 
 \end{equation} 
 and acquires meaning of an ordinary potential parameter. 
 
 \noindent 
 According to the nomenclature in \cite{Khare}, the singular interaction, 
 $V_{ScI}(\theta)$ in (\ref{ScfI}) is known under the name of the trigonometric Scarf potential,  abbreviated, `` Scarf I''. 
 This nomenclature, though strictly speaking at variance with Scarf's original work \cite{Scarf}, 
 where a $\csc^2$ singular potential has been considered, pays tribute to the $\mbox{sech}^2$ interaction, 
 commonly named  in the literature as hyperbolic Scarf potential, 
 to which it relates by a complexification of the argument directly and without the need of introducing an additional shift by $\pi/2$. 
 Instead, the $\csc^2 $ potential,   
 according to same nomenclature,  is referred to as P\"oschl-Teller potential with emphasis on \cite{PoTe} where 
 a combined  $\csc^2 +\sec^2 $ interaction has been introduced for the first time. Admittedly, other nomenclatures exist 
 \cite{Dut} according to which $\csc^2$ is referred to as trigonometric  Scarf potential. 
 As it will become clear in due course  shortly below,  after eq.~(\ref{ALF_JC}), the nomenclature does not affect the results.\\ 
 
 \noindent 
 The singular ${\mathcal H}(\theta)$ eigenvalue problem is known to be exactly solvable and has been well elaborated in the literature. 
 The general method consists in finding a suited point-canonical transformation that reduces the Schr\"odinger equation to the 
 hyper geometric differential equation, known to have three regular singular points \cite{Whittaker}, namely at $x=0$, $x=1$, and at infinity. 
 At regular points, the solutions can be expanded in Frobenius series which in the case under consideration happen to terminate and give rise to Jacobi polynomials. This method has been pioneered by Stevenson as early as 1941 in \cite{Stevenson}. 
 Alternatively, such equations can also be handled by the so called factorization method underlying modern super-symmetric 
 quantum mechanics (see \cite{Khare} for details), or by employing the recently elaborated Nikiforov-Uvarov method \cite{Suprami}. 
 
 The solutions to (\ref{Pert_Schr}) are given, among others, in  \cite{Dut}. Upon equipping the wave functions  with quantum 
 numbers according to, 
 \begin{equation} 
 U(\theta)\to U_t^{|M|} (\theta), \quad  \epsilon\to  \epsilon_t, 
 \end{equation} 
 they are expressed as, 
 \begin{eqnarray} 
 U_t^{|M|} (\theta)={\mathbf W}(\theta)\cos ^{|M|+\frac{1}{2} }\theta P_n^{|M|-b , |M|+b }(\sin\theta),&&\nonumber\\ 
 {\mathbf W}(\theta)=\left(\frac{1+\sin \theta }{1-\sin \theta }\right)^{\frac{b}{2}},\quad 
 \psi_t^M(\theta,\varphi)=U_t^{|M|}(\theta)e^{iM\varphi},&& 
 \label{wafu_Sd}\\ 
 \epsilon_t = t(t+1),\quad   t=|M|+n, \quad  |M|\in \left[0,t \right],\,\, t=0,1,2,...,&& 
 \label{ScarfI_Schr} 
 \end{eqnarray} 
 and with $n$ standing for the degree of the Jacobi polynomial, $P_n^{\alpha,\beta}(\sin\theta)$ \cite{Dennery}. 
 Setting $b=0$ recovers (modulo the $\sqrt{\cos\theta}$ factor in (\ref{var_chng})) the solution of the free 
 rigid rotator by equivalently re-expressing the spherical harmonics  in terms of the Jacobi polynomials according to, 
 \begin{eqnarray} 
 \frac{U_t^{|M|} (\theta)|_{b=0}}{\sqrt{\cos\theta}}&=& P_t^{M}(\sin\theta)= 
 \cos^{|M|}\theta P_n^{|M| , |M| }(\sin\theta), \quad n=t-|M|. 
 \label{ALF_JC} 
 \end{eqnarray} 
 The $b=0$ case corresponds to a Schr\"odinger equation with the $\sec^2\theta$ potential $V_1(\theta)$ in (\ref{IP_5}), 
 while $b\not=0$ corresponds to (\ref{ScfI}), same as $V_2(\theta)$ in (\ref{IP_5}). 
 As long as the energy in (\ref{ScarfI_Schr}) is $b$ independent, the isospectrality of the potentials in (\ref{IP_5}) is revealed. 
 Notice that because the polar angle in (\ref{L2ang}) has been read off from the negative $y$-axis instead as usual from the North Pole, 
 the argument in the associated Legendre functions is  $\cos\left(\frac{\pi}{2}-\theta \right)=\sin\theta$ instead of the usual 
 $\cos\theta$. In reality, we are dealing with Jacobi Polynomials of an argument $x$ varying in the interval, $-1\leq x\leq 1$, and 
 both choices for $x$, $x=\sin\theta$, with $\theta\in \left[ -\pi/2, +\pi/2 \right]$, or, 
 $x=-\cos \theta$ with $\theta =\in \left[ 0,\pi \right]$ are equally valid. 
 
 Accordingly, the parities of the wave functions, to be studied in the next section, are obtained from $\theta \to -\theta$ and not from $\theta \to \pi - \theta$. 
 Finally, the $M$ values are  restricted to natural numbers, satisfying the conditions, 
 \begin{equation} 
 |M|-b+1>0, \quad |M|+b +1>0, 
 \label{b_conditions} 
 \end{equation} 
 as required to ensure the  known orthogonality of the Jacobi polynomials on the interval $-1\leq\sin\theta\leq +1$. 
 As far as the $|M|$ values are integer, this restriction implies $|b|<1$ for $|M|=0$.\\ 
 
 \noindent 
 Comparison of (\ref{ScarfI_Schr}) to (\ref{Gl3})-(\ref{Gl4}), with both $t$ and $J$ taking same non-negative integer values, 
 reveals the indistinguishability of the energy spectra of the free rigid rotator and 
 the one perturbed by the trigonometric Scarf potential in the equations, (\ref{Pert_Sph})--(\ref{ScarfI_Schr}). 
 However the Scarf I wave functions in (\ref{wafu_Sd}) are distinct from the spherical harmonics in (\ref{Ylm}), as it should be, given 
 the non-commutativity of the squared angular momentum operator with this very potential, i.e. 
 $\left[ {\mathbf J}^2(\theta,\varphi),V_{ScI}({ \theta })\right]\not=0$. 
 There is indeed no theorem which prevents  two non-commuting operators of having  equal eigenvalues. 
 The operators non-commutativity only prohibits  their simultaneous diagonalizing in same function basis. 
 \begin{quote} 
 Our findings on the exotic rotational bands parallel at the quantum level the non-equivalent isospectral classical drums 
 \cite{MKac} mentioned in the introduction. 
 \end{quote} 
 The next section is devoted to  the loss of parity of the wave functions of the perturbed rigid rotator under investigation. 
 
 \section{Parityless states and non-vanishing electric dipole moments} 
 
 In the present section we study the relationship between the wave functions of the free and perturbed rotators. 
 We begin with the simplest case of the ground state, $t=0$, implying, $n=M=0$. The corresponding wave function is read off from (\ref{wafu_Sd}) as 
 \begin{eqnarray} 
 \psi_0^0(\theta,\varphi)&=&U_0^0(\theta) e^{i0\varphi}, \nonumber\\ 
 U_0^0(\theta)&=&\sqrt{\cos \theta}{\mathbf W}(\theta) =\sqrt{\cos \theta } (1+\sin\theta)^{\frac{b}{2}}(1-\sin\theta)^{-\frac{b}{2} }. 
 \label{waffugst} 
 \end{eqnarray} 
 As a technical detail, we wish to notice  that the ground state wave function, $\psi_0^0(\theta, \varphi)$, acts as 
 the eigenfunction to an $so(3)$ algebra Casimir invariant in a representation non-equivalent to the canonical one 
 and defined by  the following similarity transformation of the canonical squared angular momentum operator, ${\mathbf J}^2$, in 
 (\ref{L2ang}): 
 \begin{eqnarray} 
 {\widetilde {\mathbf J}}^2(\theta,\varphi)&=&\cos^{\frac{1}{2}}\theta {\mathbf W}(\theta){\mathbf J}^2 (\theta,\varphi) 
 {\mathbf W}^{-1}(\theta)\cos^{-\frac{1}{2}}\theta,\nonumber\\ 
 {\widetilde {\mathbf J}}^2(\theta,\varphi){\widetilde Y}_J^M(\theta,\varphi)&=& J(J+1){\widetilde Y}_J^M(\theta,\varphi),\nonumber\\ 
 {\widetilde Y}_J^M(\theta,\varphi)&=&\sqrt{\cos\theta}{\mathbf W}(\theta)Y_J^M(\theta,\varphi). 
 \label{Ffunction} 
 \end{eqnarray} 
 Indeed, the ${\widetilde {\mathbf J}}^2(\theta,\varphi)$ ground state, ${\widetilde Y}_0^0(\theta,\varphi)$, satisfies 
 \begin{eqnarray} 
 {\widetilde {\mathbf J}}^2(\theta,\varphi){\widetilde Y}_0^0(\theta,\varphi)=0, 
 \label{gstsym} 
 \end{eqnarray} 
 whose obvious solution, 
 \begin{eqnarray} 
 {\widetilde Y}_0^0(\theta,\varphi)=\cos^{\frac{1}{2}} \theta {\mathbf W}(\theta)Y_0^0(\theta,\varphi),\quad 
 Y_0^0(\theta,\varphi)=\sqrt{\frac{1}{4\pi}}, 
 \label{rscld_Y} 
 \end{eqnarray} 
 coincides  (modulo multiplicative constants) with $\psi_0^0(\theta,\varphi)$ in (\ref{waffugst}). 
 The ${\widetilde {\mathbf J}}^2$ eigenfunctions, ${\widetilde Y}_J^M(\theta,\varphi)$, are obtained by rescaling the ordinary spherical harmonics, $Y_J^M(\theta,\varphi)$, by the transformation function, $\sqrt{\cos\theta}{\mathbf W}(\theta)$, 
 and will be termed to as ``rescaled harmonics''. 
 However, due to the unspecified parity of the  function, ${\mathbf W}(\theta)$ in (\ref{Ffunction}), which is transformed by reflection into its inverse, ${\mathbf W}(-\theta)\to {\mathbf W}^{-1}(\theta)$, rather than into itself up to a sign, 
 all the rescaled harmonics are of unspecified parity. 
 {}Related  examples  can be found in \cite{AlHassid}.\\ 
 
 \noindent 
 Our next example concerns the  first excited state, corresponding to $t=1$, and to the maximal allowed $|M|$ value of $|M|=1$, 
 \textit{i.e.} $n=0$. 
 One finds, 
 \begin{eqnarray} 
 \psi_1^1(\theta,\varphi)&=&U_1^1(\theta)e^{i\varphi}={\mathbf W}(\theta)\cos^{\frac{3}{2}}\theta P_0^{-b,+b}(\sin\theta)e^{i\varphi}= 
 {\mathbf W}(\theta)\cos^{\frac{3}{2}} \theta e^{i\varphi}\nonumber\\ 
 &=& \sqrt{\cos\theta}{\mathbf W}(\theta)P_1^1(\sin\theta)e^{i\varphi}= \widetilde{Y}_1^1(\theta,\varphi), 
 \label{t1m1} 
 \end{eqnarray} 
 and again an eigenfunction to the ${\widetilde {\mathbf J}}^2$  $so(3)$ Casimir invariant in (\ref{Ffunction}), 
 \begin{equation} 
 {\widetilde {\mathbf J}}^2(\theta,\varphi){\widetilde Y}_1^1(\theta,\varphi)=2{\widetilde Y}_1^1(\theta,\varphi). 
 \label{test_gst} 
 \end{equation} 
 In general, it can be shown that all wave functions of maximal  $|M|=t$ values (they include the ground sate) behave as 
 ${\widetilde {\mathbf J}}^2(\theta,\varphi)$ 
 eigenfunctions and are characterized by a single angular momentum value, same that defines their energy. 
 {}For all these states one is allowed to maintain the $t$ label as a number conserved under transformations of the $so(3)$ algebra in the representation defined in (\ref{gstsym}), despite the loss of the parity specification explained after the equation (\ref{rscld_Y}) 
 above. \\ 
 
\begin{figure}[H]
\centering
\begin{tabular}{cc}
\resizebox{5cm}{!}{\includegraphics{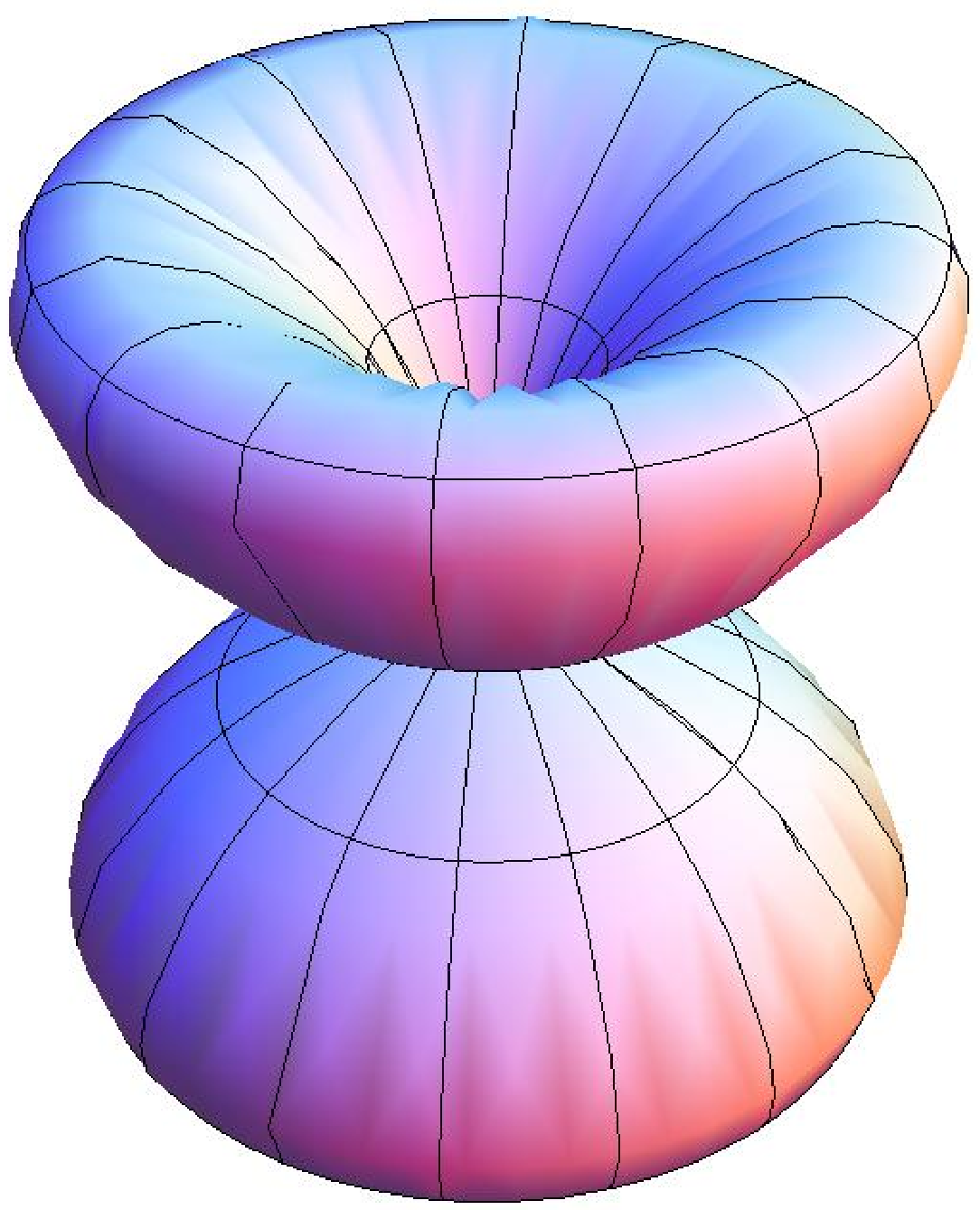}} & \resizebox{5cm}{!}{\includegraphics{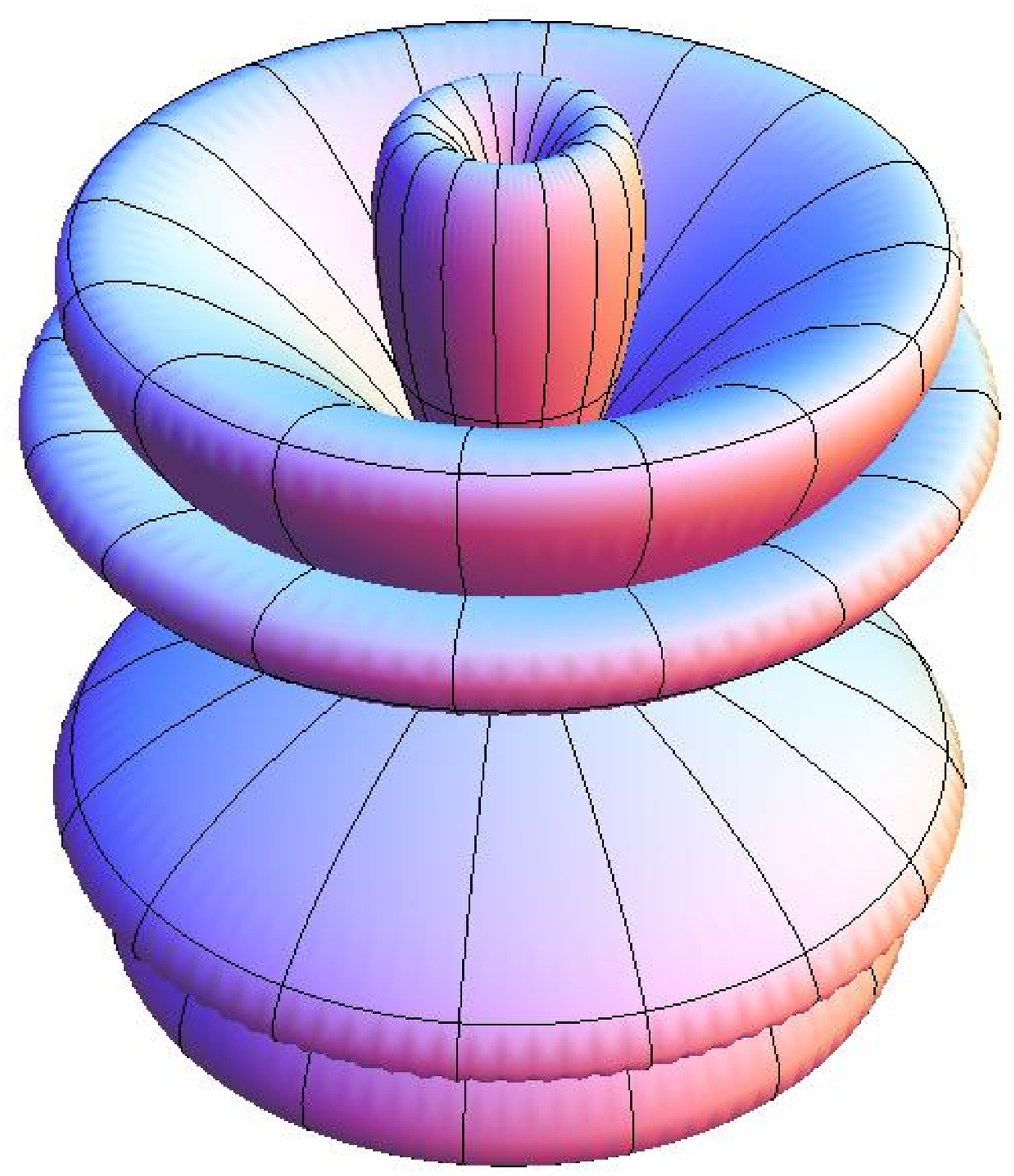}} \\
(a) $|Y_{2}^{1}\left(\theta,\varphi\right))|^2$ & (b) $|\psi_{2}^{1}\left(\theta,\varphi\right)|^2$
\end{tabular}
\caption{Illustration of how equal rotational energies (frequencies) can correspond to non-equivivalent shapes of the
 probability density distributions.
{}For the canonical rigid rotor, a probability density distribution of the release to the ground state 
of the  energy,  $E_J=E_t=6{\mathcal B}$, of the second excited level, 
is given in fig.~2(a), while  that corresponding to the perturbed rotator is displayed in fig.~2(b). 
The squared canonical rigid rotator wave function in fig.~2(a) is single $J$- valued,  
while the perturbed one  in fig.~2(b) contains  a $\Delta J=1$  mixture  according to (\ref{Bong}).
The potential  parameter $b$ has been given the generic continuous value of, $b=0.45$.
The figure visualizes the statement that one can not "hear" the shape of the radiation patterns from the spectral frequencies.}
\label{Fig:2}       
\end{figure}

 \noindent 
 Next we consider  consider the case of 
 $t=2$ and $|M|=1$, which corresponds to $n=1$. We again  express the relevant wave function $\psi_2^1(\theta,\varphi)$   
 in terms of associated Legendre functions, $P_J^{|M|}(\sin\theta)$. In so doing, one finds 
 \begin{eqnarray} 
 \psi_2^1(\theta,\varphi)&=&\sqrt{\cos\theta}{\mathbf W}(\theta)\cos\theta P_1^{1-b,1+b}(\sin\theta)e^{i\varphi}\nonumber\\ 
 &=&\sqrt{\cos\theta}{\mathbf W}(\theta)\cos\theta\,  \left[-b +2\sin\theta \right]\, e^{i\varphi}\nonumber\\ 
 &=&\sqrt{\cos\theta}{\mathbf W}(\theta)\, \left[ -b\cos\theta +2\cos\theta\sin\theta \right]\, e^{i\varphi}\nonumber\\ 
 &=&\sqrt{\cos\theta}{\mathbf W}(\theta)\, \left[ -bP_1^1(\sin\theta)e^{i\varphi} +\frac{2}{3}P_2^1(\sin\theta) e^{i\varphi}\right]\nonumber\\ 
 &=&\sqrt{\cos\theta}{\mathbf W}(\theta)\, \left[ -b { Y}_1^1(\theta,\varphi) +\frac{2}{3}{ Y}_2^1(\theta,\varphi)\right]\nonumber\\ 
 &=&-b{\widetilde Y}_1^1(\theta,\varphi) +\frac{2}{3}{\widetilde Y}_2^1(\theta,\varphi). 
 \label{Bong} 
 \end{eqnarray} 
 As long as the parity of the spherical harmonics  is $(-1)^J$, the functions $Y_1^1(\theta,\varphi)$, and $Y_2^1(\theta,\varphi)$ 
 are of opposite parities, and $\psi_{2}^1(\theta,\varphi )$ is built on top of a parity-mixed state. Moreover, the mixture of 
 the rescaled harmonics in (\ref{Bong}) no longer behaves as 
 ${\widetilde {\mathbf J}}^2$ eigenstate, and the $t=2$ label in this case can not be interpreted 
 as a quantum number conserved under transformations of the $so(3)$ algebra in the representation of (\ref{gstsym}). 
 Decompositions of the art presented in (\ref{Bong}) can be performed for all the higher  $t$ values. 
 They  conserve the $|M|$ label and mix (rescaled) spherical harmonics with $J =|M|, ..., t$, 
 \begin{eqnarray} 
 \psi_t^{M}(\theta, \varphi)&=& \sum_{J=|M|}^{J=t}c_J {\widetilde Y}_J^{M}(\theta, \varphi). 
 \label{gen_wafu} 
 \end{eqnarray} 
 As a reminder, the  ${\widetilde Y}_J^{M}(\theta, \varphi)$ functions, defined in (\ref{Ffunction}), 
 are always of unspecified parity. 
 It is obvious that the electric dipole moment, whose selections rules are $\Delta M=0$, and $\Delta J=1$, 
 will have a non-vanishing expectation value 
 in the $\psi_2^1(\theta,\varphi)$ state due to the allowed  $Y_2^1(\theta,\varphi)\to Y_1^1(\theta,\varphi)$ transition, on the one side, 
 and more generally, due to the parityless nature of the ${\mathbf W}(\theta)$ function, on the other side. 
 We have checked that also the general  integrals 
 \begin{eqnarray} 
 d_e&=&\int \left[\psi^M_t(\theta,\varphi)\right]^\ast \sin\theta \psi^M_t(\theta,\varphi)\sin\theta {\mathrm d}\theta{ \mathrm d}\varphi\not=0, 
 \label{edm} 
 \end{eqnarray} 
 are non-vanishing for $b$ values satisfying the conditions in (\ref{b_conditions}). 
 We recall that the argument of the associated Legendre functions in these spherical harmonics is 
 $\sin\theta $, versus $\cos\theta$ in the standard ones. 
 Finally a comment on the symmetry properties of the wave functions of the perturbed motion is in order. 
 These can vary and are determined besides by the $|M|$ value, as explained in the paragraph after eq.~(\ref{test_gst}), 
 also  by the values of the  $b$ parameter. While for continuous $b$ values the wave functions with $|M|\not=t$   
 do not behave as ${\widetilde {\mathbf J }}^2$ eigenfunctions,  for regular integer, or half-integer $b$ values, 
 call them $b=M^\prime$, it can be shown that the Jacobi polynomials times $\cos^{|M|}\theta$ would behave as Wigner's small $d^t_{M M^\prime}$ functions \cite{Levai} 
 and the rotational symmetry is respected. We here emphasize on continuous $b$ values. 
 Comparison of the ordinary spherical harmonics, the wave functions of the free rigid rotator, 
 to the Schr\"odinger wave functions of the Scarf I- perturbed rotator, is presented in fig. \ref{Fig:2}. 
 
 \section{Conclusions} 
 
 In conclusion, pure rotational frequency patterns  do not necessarily 
 imply equality of the Hamiltonian 
 to that  of the canonical rigid rotator. 

 {}For discrete values of the potential parameter $b$,  such patterns occur 
 because the isospectral potentials in (\ref{IP_5}) 
 are related through a similarity transformation of the type given in 
 (\ref{IIP_2b})-(\ref{IP_2c}) with the place of ${\mathbf F}$ 
 being  taken by the $\cos^{\frac{1}{2}}\theta {\mathbf W}(\theta)$ factor 
 in the wave function in (\ref{wafu_Sd}). We discussed this issue after 
 eq.~(\ref{edm}).  Stated differently, the Schr\"odinger Hamiltonian with 
 the trigonometric Scarf potential for such parameter values is identical 
 to the  similarity transformed 
 squared angular momentum operator, $\widetilde{\mathbf 
 J}^2=\cos^{\frac{1}{2}}\theta {\mathbf W}(\theta){\mathbf J}^2 
 \cos^{-\frac{1}{2}}\theta {\mathbf W}^{-1}(\theta)$, as explained in 
 (\ref{Ffunction}). Same happens when the magnetic quantum number 
 $|M|$ takes its maximal value. In all these cases the reason behind the 
 rotational patterns of the perturbed motion 
  is still the $so(3)$ symmetry of the Hamiltonian, though in a 
 representation distinct from the canonical one. 
 However, for continuous $b$ values the above considerations are no longer 
 valid and 
 the reason behind the rotational spectrum is not obvious. In order to get 
 an insight into the symmetry properties of the 
 perturbed Hamiltonian for this case one has to follow the standard 
 procedure and find the complete set of operators with which it is 
 commuting. In calculating the commutators of these operators, a clue on a 
 possible Lie algebra might be obtained \cite{we}. 
 
 In all cases, the  wave functions of the rigid rotator, perturbed by the 
 trigonometric Scarf potential in (\ref{ScfI}), 
 despite of the rotational multiplicity of the states in the levels, are in 
 first place always of unspecified parities. 
 In consequence, a sort of exotics is produced in so far as the states in 
 the rotational bands acquire non-vanishing electric 
 dipole moments. 
 The symmetry behind the emerging exotic rotational spectrum for continuous 
 $b$ values 
 is not very clear so far. We conclude on the possibility for existence of 
 exotic rotational bands containing parityless states in 
 two-body systems of a possibly new class,  be them di-atomic-, 
 di-molecular, quarkish,  etc.  and  whose states 
 carry non-vanishing electric dipole expectation values. 

\section*{Acknowledgement}
We thank an anonymous referee for constructive criticism and detailed valuable advice on several mathematical aspects of the presentation.

 \end{document}